\documentclass[prl,twocolumn,amsfonts,superscriptaddress]{revtex4}
\usepackage{graphicx}
\usepackage{color}
\usepackage{portland}

\begin{document}

\title{Quantitative imaging of flux vortices in type-II superconductor MgB$_2$ using Cryo-Lorentz Transmission Electron Microscopy}

\author{M.J.G. Cottet}
\affiliation{Laboratory for Ultrafast Microscopy and Electron Scattering, ICMP, Ecole Polytechnique F\'{e}d\'{e}rale de Lausanne, CH-1015 Lausanne, Switzerland}
\author{M. Cantoni}
\affiliation{Centre Interdisciplinaire de Microscopie Electronique, CIME, Ecole Polytechnique F\'{e}d\'{e}rale de Lausanne, CH-1015 Lausanne, Switzerland}
\author{B. Mansart}
\affiliation{Laboratory for Ultrafast Microscopy and Electron Scattering, ICMP, Ecole Polytechnique F\'{e}d\'{e}rale de Lausanne, CH-1015 Lausanne, Switzerland}
\author{D.T.L. Alexander}
\affiliation{Centre Interdisciplinaire de Microscopie Electronique, CIME, Ecole Polytechnique F\'{e}d\'{e}rale de Lausanne, CH-1015 Lausanne, Switzerland}
\author{C. H\'{e}bert}
\affiliation{Centre Interdisciplinaire de Microscopie Electronique, CIME, Ecole Polytechnique F\'{e}d\'{e}rale de Lausanne, CH-1015 Lausanne, Switzerland}
\author{N.D. Zhigadlo}
\affiliation{Laboratory of Solid State Physics, ETH Zurich, CH-8057 Zurich, Switzerland}
\author{J. Karpinski}
\affiliation{Laboratory of Solid State Physics, ETH Zurich, CH-8057 Zurich, Switzerland}
\affiliation{Laboratory of Nanostructures and Novel Electronic Materials, ICMP, Ecole Polytechnique F\'{e}d\'{e}rale de Lausanne, CH-1015 Lausanne, Switzerland}
\author{F. Carbone}
\affiliation{Laboratory for Ultrafast Microscopy and Electron Scattering, ICMP, Ecole Polytechnique F\'{e}d\'{e}rale de Lausanne, CH-1015 Lausanne, Switzerland}

%\pacs{}

\begin{abstract}

Imaging of flux vortices in high quality MgB$_2$ single crystals has been successfully performed in a commercial Field Emission Gun-based Transmission Electron Microscope. In Cryo-Lorentz Microscopy, the sample quality and the vortex lattice can be monitored simultaneously, allowing one to relate microscopically the surface quality and the vortex dynamics. Such a vortex motion ultimately determines the flow resistivity, $\rho_{f}$, the knowledge of which is indispensable for practical applications such as superconducting magnets or wires for Magnetic Resonance Imaging. The observed patterns have been analyzed and compared with other studies by Cryo-Lorentz Microscopy or Bitter decoration. We find that the vortex lattice arrangement depends strongly on the surface quality obtained during the specimen preparation, and tends to form an hexagonal Abrikosov lattice at a relatively low magnetic field. Stripes or gossamer-like patterns, recently suggested as potential signatures of an unconventional behavior of MgB$_2$, were not observed.

\end{abstract}


\maketitle

\begin{figure}[ht]
\includegraphics[width=1\linewidth,clip=true]{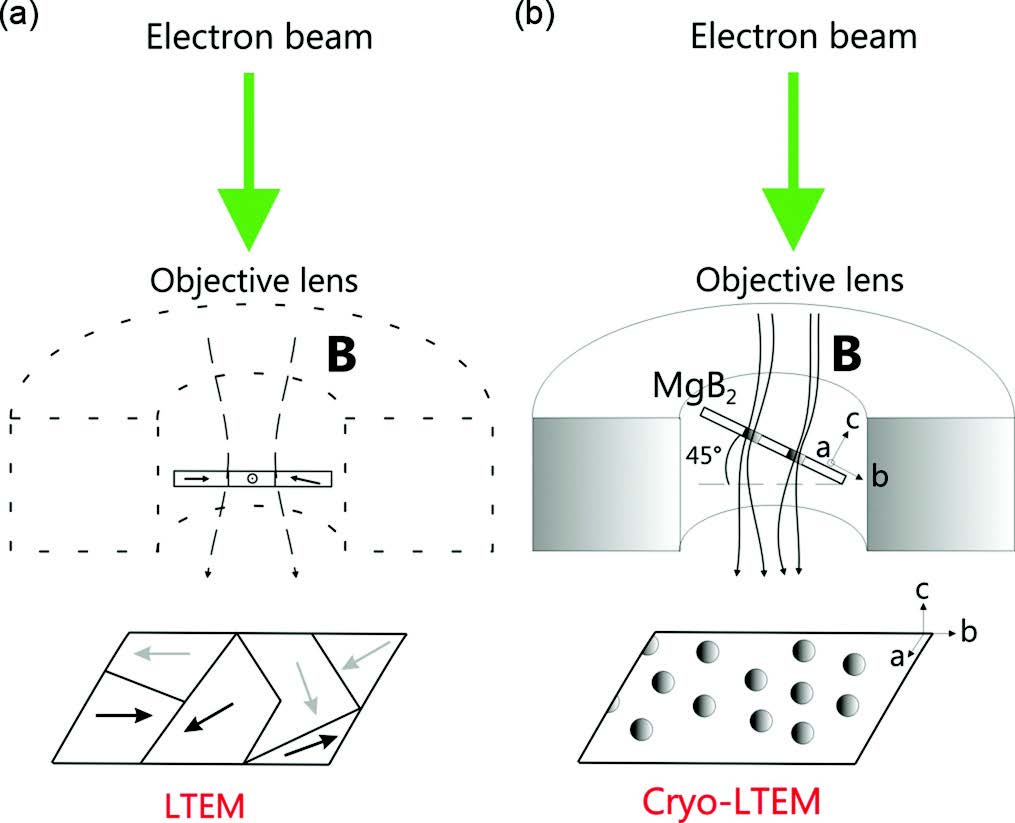}
\caption{Lorentz microscopy or phase contrast microscopy. A) Example of Lorentz Transmisson Electron Microscopy (LTEM) on a magnetic compound; there is a small residual magnetic field near the specimen, the objective lens is off. B) Cryo-Lorentz Transmission Electron Microscopy on a superconductor (Cryo-LTEM), the objective lens is weakly activated to induce a magnetic field near the specimen. The electrons are deflected by the magnetic fields of the vortices giving rise to black and white features as shown on the image plane.}
\label{lorentz}
\end{figure}

In superconductors, identifying the relationship between the transverse force felt by the triangular array of vortices in response to a transport current and the pinning force resulting from defects or inhomogeneities is of capital importance. This is because flux motion induces a longitudinal resistive voltage, which is a source of energy dissipation and ultimately hinders a material's performance in applications \cite{Matsuda}. Indeed, in type-II superconductors, the superconducting state is not completely destroyed when an external magnetic field exceeds the lower critical one, $\textit{H}_{c1}$. The external field partially penetrates into the material in the form of midget microscopic filaments called vortices. As a first approximation (Bardeen-Stephen model \cite{Bardeen}), the core of each vortex, where superconductivity is supressed, is modeled by a cylinder with a radius given by the coherence length $\xi$, and is surrounded by circling supercurrents over a distance corresponding to the London penetration depth $\lambda$. It is assumed that the core of each vortex is a conventional metallic state inside of which the energy dissipation is dominated by impurity scattering \cite{Kim}. Each vortex carries a magnetic flux equal to $\Phi_0$ = $\textit{h}$/(2$\textit{e})$, where \textit{h} is the Planck constant and \textit{e} the elementary charge \cite{Deaver}. Due to their mutual repulsion, in a defect free superconductor, vortices tend to form a 2D close-packed triangular array surrounded by an hexagonal pattern of other vortices, called the Abrikosov lattice \cite{Abrikosov}. Cryo-Lorentz Transmission Electron Microscopy (Cryo-LTEM) allows the direct observation of quantized flux lines and so is a key technique for understanding the flux flow resistivity associated with the viscous motion of the vortices \cite{Shibata}. It is with the out-of focus imaging of Lorenz mode that individual flux quanta can be imaged, as well as superconducting or magnetic domains (Fig.~\ref{lorentz}, Fresnel mode) \cite{Harada}. Specifically, flux lines can be imaged thanks to the deflection imparted to the electrons by the magnetic flux associated with each vortex. In this paper, we investigate the formation and the dynamics of flux vortices in MgB$_2$ single crystals using Cryo-LTEM with a defocus of $\approx15$ mm \cite{Chapman}.

MgB$_2$ is a type-II superconductor with a T$_c$ = 39 K. It crystallizes in an hexagonal structure consisting of alternating honeycomb layers of B and hexagonal layers of Mg atoms, with \textit{a} = \textit{b} = 3.1432 $\pm$ 0.0315 $\buildrel _{\circ} \over {\mathrm{A}}$ and \textit{c} = 3.5193 $\pm$ 0.0323 $\buildrel _{\circ} \over {\mathrm{A}}$ \cite{Budko}. In the low field limit, the coherence length in the $ab$ plane has been found as $\xi_{ab}$ = 5.5 nm \cite{LoudonPRB} and $\lambda_{ab}$ lies between 110 - 130 nm \cite{Manzano} ($\xi_c$ = 51 nm and $\lambda_c$ = 33.6 nm \cite{Moshchalkov1}. The lower critical field $\textit{H}_{c1}$ in the \textit{c}-direction is around 250 mT and the upper critical field $\textit{H}_{c2}$ is lower than 3 T at 5 K \cite{Caplin}. MgB$_2$ is a two bands system with two different gap amplitudes: $\Delta_\sigma$ = 7.1 $\pm$ 0.1 meV following a BCS-like temperature dependence, and $\Delta_\pi$ = 2.80 $\pm$ 0.05 meV showing deviations at T $\geq$ 25 K as shown by Gonnelli \textit{et al}. \cite{Gonnelli}.

Recently, Gurevich \textit{et al}. \cite{Gurevich} have demonstrated the tunability of $\textit{H}_{c2}$ for different MgB$_2$ samples when alloyed with nonmagnetic impurities; $\textit{H}_{c2}$ values well above those of some competitive non-cuprate superconductors were obtained, thereby increasing the potential applications of this compound. According to the value of the Ginzburg Landau parameter in the \textit{ab}-palne reported by Caplin \textit{et al.} ($\kappa \cong$ 4 at 5 K),  this material is far into the limit of type-II superconductors \cite{Caplin}. However, this classification has been recently questioned by the proposition of a novel superconducting state in-between type-I and type-II which could stabilize in an unconventional vortex pattern formed by stripes and gossamer-like shapes (vortex clusters emerging at low magnetic field). The observation of such a vortex pattern has been reported by Moshchalkov \textit{et al.} \cite{Moshchalkov1} and Nishio \textit{et al.} \cite{Nishio} at very low fields, while, more recently, Cryo-Lorentz TEM experiments have instead shown a distorted square lattice arrangement surrounded by vortex lines resulting from the specimen preparation \cite{Loudon}, different from a conventional and stable triangular pattern \cite{Fischer, Cribier, Kleiner}.

\begin{figure}[ht]
\includegraphics[width=1\linewidth,clip=true]{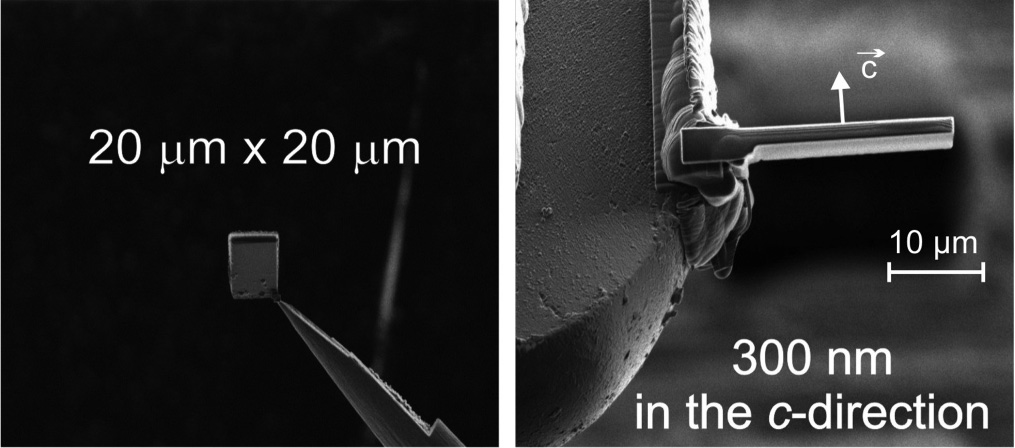}
\caption{Images of the specimen during the preparation process. The sample is pre-tilted at 45$^\circ$ when attached to a dedicated TEM grid, the sample is electron transparent in the \textsl{c}-direction.}
\label{FIB}
\end{figure}

In this article, we perform new Cryo-LTEM experiments on MgB$_2$ and confirm that its distribution of flux quanta follows a conventional type-II behavior. Vortices tend to form a stable triangular pattern at relatively low magnetic fields due to the symmetrical surrounding resulting from short range repulsion \cite{Fletter}. Stripes or gossamer-like patterns are not observed, although distortions of the expected triangular arrangement are visible close to the sample edge. This suggests that the presence of such other patterns probably results from a non-ideal surface obtained during the sample preparation process, and does not necessarily justify the assignment of an unconventional class of superconductors by means of surface sensitive techniques such as Bitter decoration \cite{Moshchalkov1, Nishio}. In fact, to be in equilibrium in a certain position, the vortex lines must have a total superfluid velocity equal to zero. This can be accomplished if each vortex of the pattern is surrounded by a symmetrical array like a triangular or a square pattern. However, the square array exhibits only an unstable equilibrium since small displacements tend to be spontaneously generated; the triangular lattice instead has the lowest free energy. This is confirmed by the solution of the Ginzburg-Landau equations of the Abrikosov type for a defect free type-II superconductor for magnetic fields below H$_{c2}$, giving the value of the parameter determining the most favorable configuration of all periodic solutions for both patterns, $\beta_A$ ($\beta_A = \left\langle \Psi^4_L \right\rangle/\left\langle \Psi^2_L \right\rangle^2$, $\beta^{tri}_{A}= 1.16$ and $\beta^{sq}_{A} = 1.18$), where $\left\langle \Psi_L \right\rangle$is a general solution to the linearized Ginzburg Landau equation \cite{Abrikosov, Kleiner}. Moreover, pinning results from spatial inhomogeneity since local variations of $\xi$, $\lambda$ or $H_{c}$ leads to a local change of the free energy per unit of vortex line, causing some vortices to be preferentially located at certain positions. To be relevant, the material defects must stay in the order of $10^{-6}$ to $10^{-5}$ cm, in order to avoid electronic scattering which limits the mean free path, $l$ \cite{Tinkham}.

MgB$_2$ single crystals were synthesised via the decomposition of MgNB$_9$, as described in Ref. \cite{Karpinski}. To be electron transparent, the specimen was prepared using a Zeiss NVision 40 Focused Ion Beam (FIB): the sample was thinned to 300 nm in the \textit{c}-direction, and pre-tilted at 45$^\circ$ when attached on a TEM grid (as shown in Fig.~\ref{FIB}). A Gatan single tilt Helium Cooling Holder was used to cool the specimen down to 5 K, and the images were acquired digitally on a Gatan 2k x 2k CCD camera mounted on a JEOL 2200FS TEM using a field-emission gun operated at 200 keV. 

\begin{widetext}

\begin{figure}[ht]
\includegraphics[width=1\linewidth,clip=true]{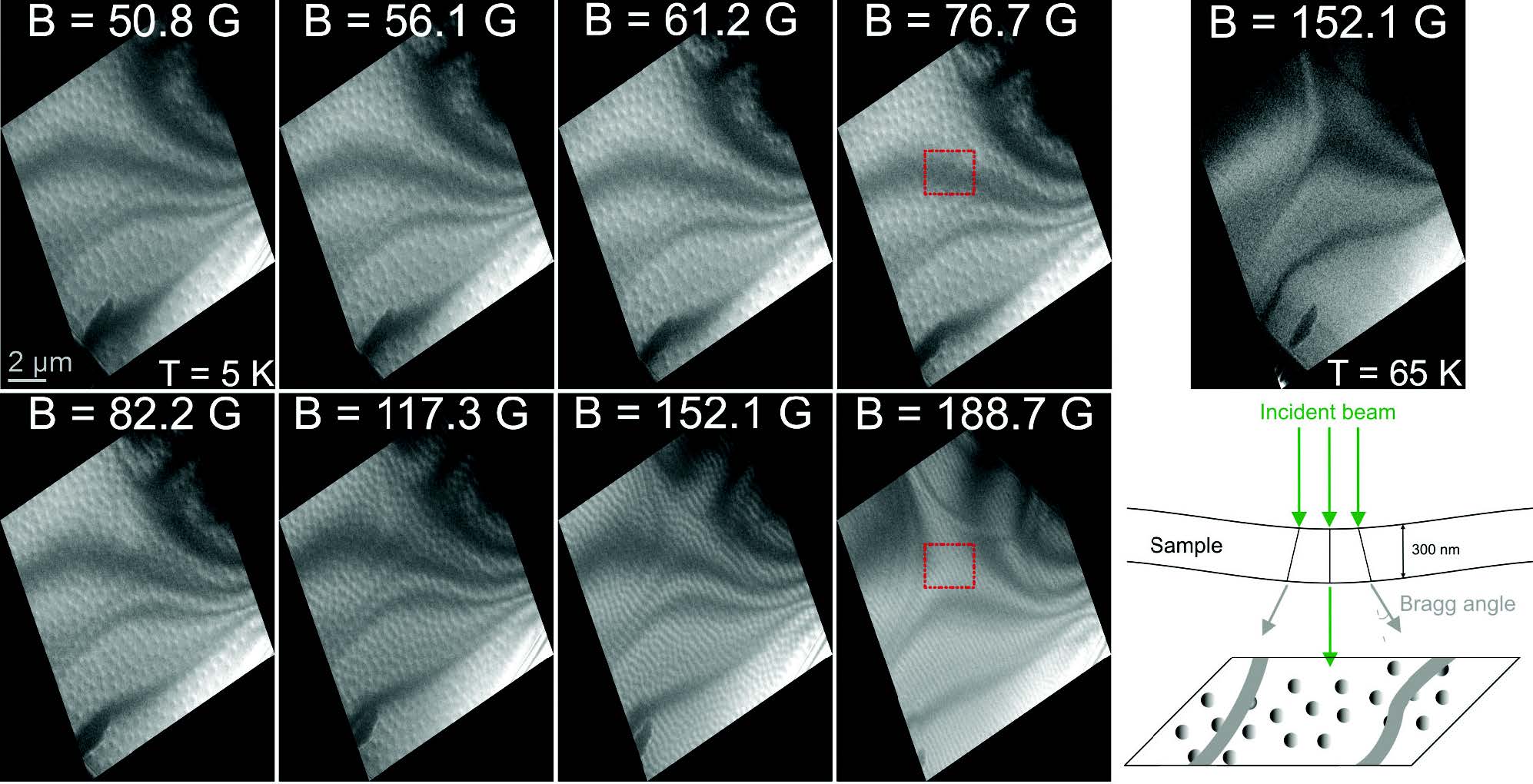}
\caption{Serie of tilt-compensated images showing flux vortices in MgB$_2$ taken at different magnetic fields (indicated on the images) and at a temperature of 5 K. The specimen is pre-tilted at 45$^\circ$ ($\pm$ 1$^\circ$). The scheme at the bottom right shows the origin of the bend contours; when the sample is locally curved the incident beam is not parallel to the \textit{hkl} plane, the Bragg condition is then fulfilled resulting in darks bands in the obtained image. The dashed rectangles show how bend contours move as a function of the induced magnetic field (temperature or small tilt difference).}
\label{vortex}
\end{figure}
\end{widetext}

In Figure~\ref{vortex}, a sequence of images showing flux vortices taken at 5 K and for different induced magnetic fields (specified on each image) is presented. The intensity of the field has been calculated via the density of vortices averaged across the region of interest. The images are tilt-compensated to correct for the apparent compression of the vortices in one direction due to the sample's configuration required to carry out Cryo-LTEM (Fig.~\ref{lorentz}. B)). The sequence begins at 50.8 G and the magnetic field is slowly increased to reach 188.7 G, where individual quanta cannot be distinguished anymore. When the temperature is increased above the superconducting transition temperature (T = 65 K) at a fixed magnetic field, no vortices are observed, as shown in the top right panel of Fig.~\ref{vortex}. At low field, vortices are formed at the edge of the specimen \cite{Olsen}, and when the magnetic flux further penetrates an ideal type-II superconductor, the force density on the vortices, $\alpha$, resulting from the Lorentz force tends to make them move \cite{Tinkham}.

\begin{equation}
\alpha = J_{ext} \times \frac{B}{c} = (curl H) \times \frac{B}{4 \pi}
\end{equation}

where $J_{ext}$ represents the externally imposed currents, leaving out the equilibrium contribution coming from the response of the specimen.

The dark bands (or areas) present in each image are called bend contours and are due to the diffraction contrast. To confirm this origin of such dark bands, it was checked that they vary in size and position when tilting the sample or changing the induced magnetic field (due to the imaging condition), as shown by the red dashed rectangles of Fig. \ref{vortex} located at the same position (the origin of bend contours is shown in the bottom-right panel of Fig.~\ref{vortex}) \cite{Marco}. As discussed later, flux vortices have poor contrast within these bend contours, impeding their analysis.

To obtain a quantitative information related to the vortex network, we calculate the sixfold bond-orientational order parameter, $\left|\Psi_6\right|$. This parameter measures the instantaneous crystalline orientation between vortices separated by a distance $\textit{r}_{ij}$ and an angle $\theta_{ij}$ with two of its direct neighbors $\textit{n}_i$. $\left|\Psi_6\right|$ = 1 for a perfect Abrikosov lattice \cite{Sow}.

\begin{equation}
\Psi_6(r_{ij},t) = \frac{1}{n_i}\sum\limits_{j=0}^{n_i} exp(6i\theta_{ij})
\end{equation}

\begin{figure}[ht]
\includegraphics[width=1\linewidth,clip=true]{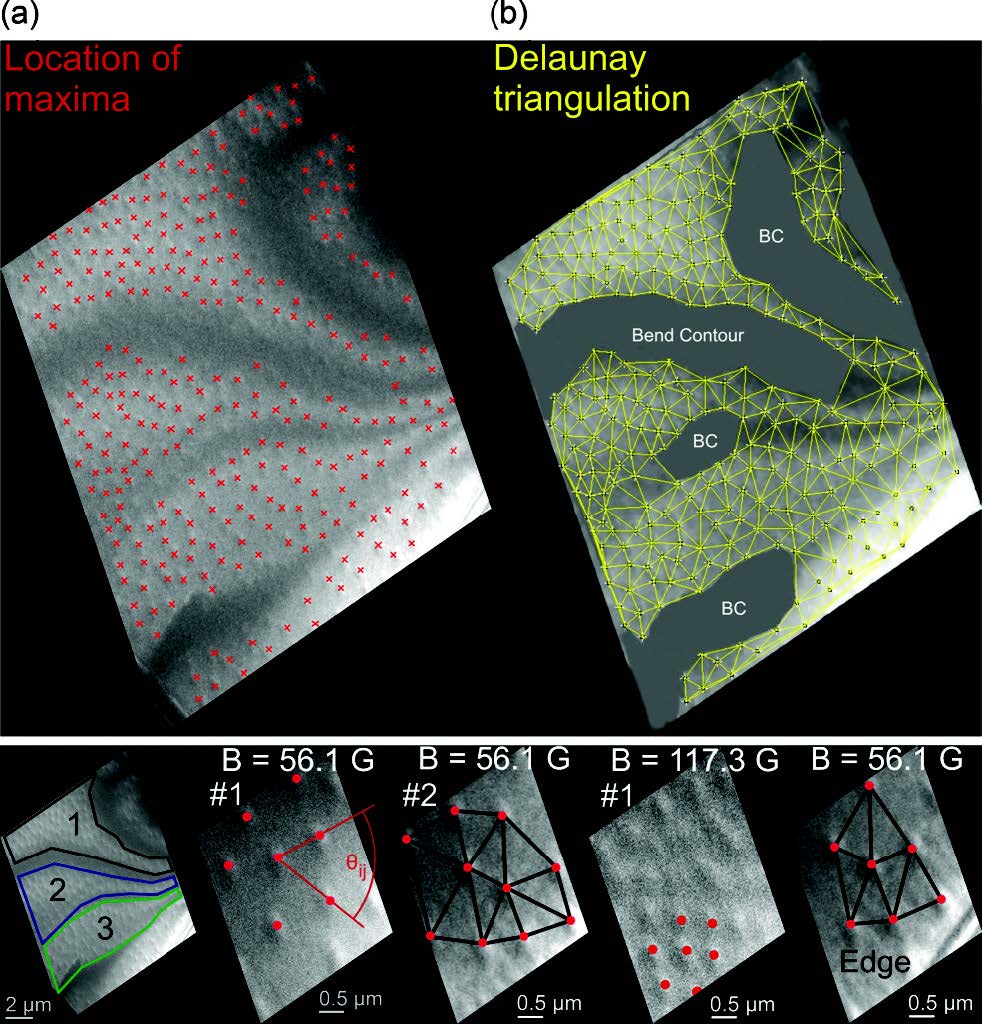}
\caption{A) Location of vortices (point location step), each red cross represents a maximum. Image taken at T = 5 K and B = 56.1 G. B) Delaunay triangulation for the set of points determined in A), each vortex is connected to its direct neighbors. The grey areas represent the bends contours (excluded for the triangulation). The bottom images (higher magnification, 4x) show examples of Delaunay triangulation of triangular arrays of vortex patterns for different induced magnetic fields (56.1 G and 117.3 G) and different areas located between the bend contours; the bottom right image shows a distorted pattern close to the sample edge.}
\label{triangulation}
\end{figure}

This parameter was calculated by using a distribution map of the vortices obtained via a triangulation algorithm, which is a useful tool for efficiently solving many topography problems \cite{Delaunay}. The Fiji open-source platform was used to carry out this triangulation, enabling fast image-processing \cite{Fiji}. The locating of vortices is accomplished by determining the local maxima in an image and creating a binary image of the same size. An example is shown in Figure~\ref{triangulation} A). The Delaunay triangulation minimizes the maximum possible circumcircle for all triangles joining a given set of points, and it maximizes the minimum angle of those triangles (Fig.~\ref{triangulation} B)) \cite{Meisner}.

\begin{widetext}

\begin{figure}[ht]
\includegraphics[width=1\linewidth,clip=true]{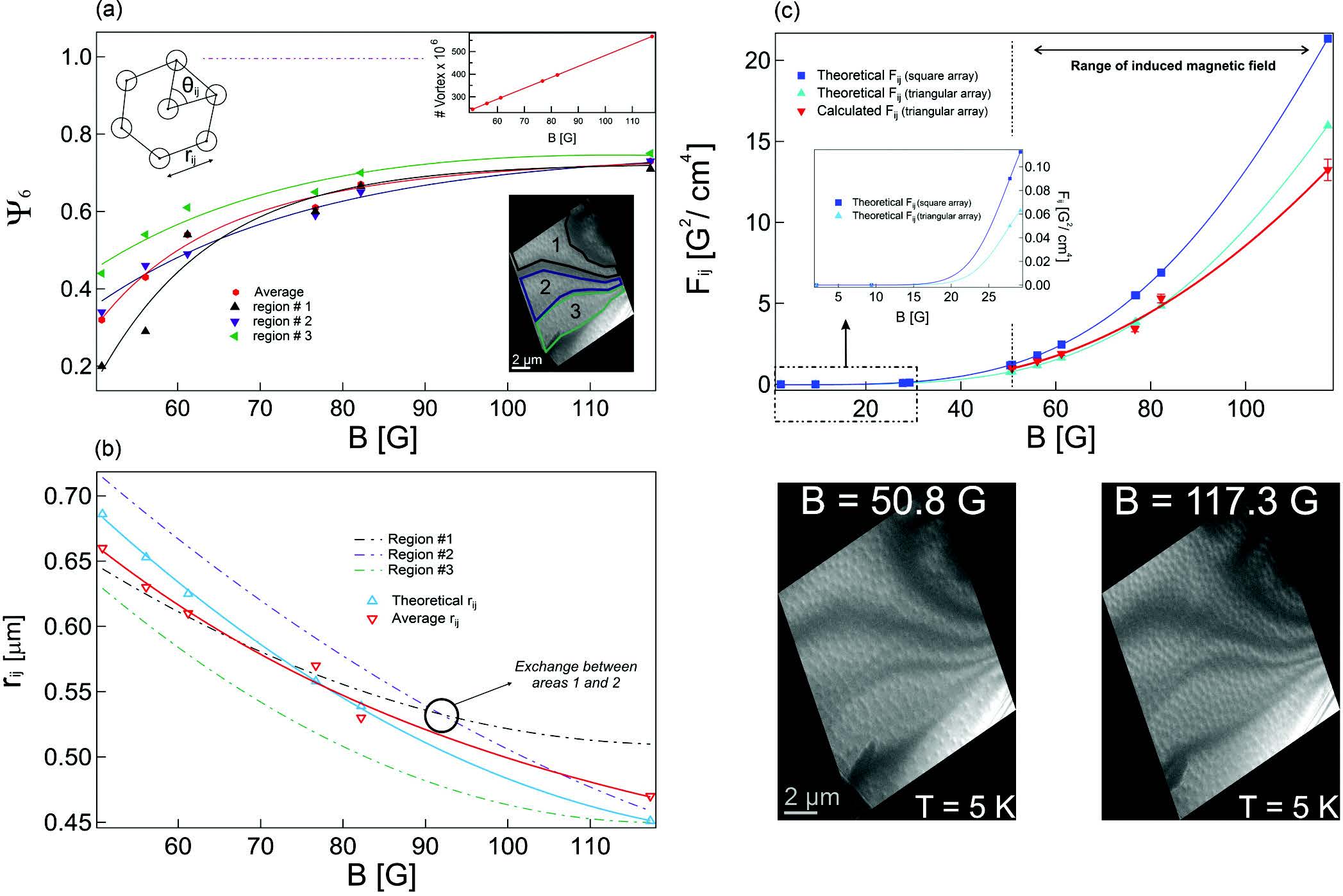}
\caption{A) Evolution of the sixfold parameter, $\left|\Psi_6\right|$, as a function of the induced magnetic field, $B$, $\left|\Psi_6\right|$ = 1 (dashed line) represents a perfect triangular array of vortices. The inset upper right graphic shows the evolution of the vortex density. B) Evolution of the average distance between vortices, $r_{ij}$, as a function of $B$ for the three regions. C) Evolution of the interaction energy, $F_{ij}$, for a pair of vortices in the low flux density regime calculated from our data; the blue and the red curves represent respectively the theoretical evolution of $F_{ij}$ for a square and a triangular array of vortices; the inset graph shows the evolution of $F_{ij}$ at very low fields for the same patterns.}
\label{six}
\end{figure}
\end{widetext}

If the sample is divided into three areas separated by the bend contours (position taken at low field, below 120 G, see Fig. \ref{vortex}), the local value of $\left|\Psi_6\right|$ can be calculated to assess the homogeneity of the magnetic field in the sample related to the surface quality; these three areas are considered as defect free, the small portion of the sample close to the edge was not analyzed due to the presence of pinned vortices and distorted patterns. An example of this procedure is given in Fig.~\ref{triangulation} where the vortex lattice in the different areas at the same magnetic field is displayed together with the corresponding Delaunay triangulation. In these areas, values of $\left|\Psi_6\right|$ equal to 0.29, 0.46 and 0.54 are found (average value corresponding respectively to the area labelled 1, 2 and 3); the same analysis was carried out throughout the specimen and for different magnetic fields obtaining a rather homogenous distribution of $\left|\Psi_6\right|$ values; the vortices are not analyzed in the dark areas of the bend contours because the signal to noise ratio does not allow the accurate determination of their locations. In Fig.~\ref{six}, the sixfold parameter, the distance between vortices and the interaction energy, $F_{ij}$, are all displayed for the three areas of the sample. In the low flux density regime ($\Phi_0 / B >> \lambda^2$), $F_{ij}$ is given by:

\begin{equation}
\sum_{i>j} F_{ij} = \left(\frac{B}{\Phi_0}\right) \frac{z}{2} \frac{\Phi^{2}_{0}}{8 \pi^2 \lambda^2_{ab}} K_0 \left(\frac{r_{ij}}{\lambda_{ab}}\right)
\end{equation}
\begin{equation}
\sum_{i>j} F_{ij} \approx \frac{B z \Phi_0}{16 \pi^2 \lambda^2_{ab}} \left(\frac{\pi \lambda_{ab}}{2 a}\right)^{1/2} e^{-r_{ij}/\lambda_{ab}}
\end{equation}

where z represents the number of neighbors (z = 6 for a triangular array and z = 4 for a square pattern); the theoretical values were obtained using $r_{ij} = d(\Phi_{0} / B)$, with $d = 1.075$ for the triangular array and $d = 1$ for the square one \cite{Tinkham}. All contributions except those coming from the nearest neighbors are neglected. According to the exponential term, for low magnetic fields, the triangular array will be lower in energy than the square pattern. The derivative of $F_{ij}$ with respect to $B$ shows that the system undergoes a second-order phase transition at $H_{c1}$, where the induced magnetic field is continuous. In other words, there is a certain range of induced magnetic fields for which only one vortex will be formed until $B$ becomes large enough to generate vortices separated by a distance $\approx \lambda$ for relatively high $\kappa$-systems ($\xi_{ab}$ = 5.5 nm and $\lambda_{ab} \cong$ 120 nm), while $B$ is discontinuous at $H_{c1}$ for low $\kappa$-systems \cite{Auer}.

A similar trend is observed in these three regions: upon increasing magnetic field, the vortex density increases and the sixfold parameter tends to approach the Abrikosov value, even though our range of analyzed magnetic field does not allow the visualization of a perfect undistorted triangular array of vortices (to reach higher $\left|\Psi_6\right|$ values, higher spatial resolution in Cryo-LTEM is needed in order to analyze the images taken at higher magnetic fields). The trend of $F_{ij}$ (Fig. \ref{six} C)) clearly confirms the stability of the triangular pattern over the square one for this regime, in which small displacements tend to grow.
When the supercurrent starts to flow the vortices start to move under the Lorentz force if they are not pinned by the so-called "pinning centers". In this experiment, the force redistributing vortices within the specimen is enough to have no vortex pinning (except at the edge), confirming the high quality of the sample surface and the close relation between the sample preparation and the observed behavior. In fact, when the pinning is weak with respect to the driving force, vortices will start to move in a steady motion with a velocity related to the viscous damping of the material, without jumping from one pinning site to another. This regime is called flux flow \cite{Solomon}.
The calculation of the sixfold parameter, as well as the average distance between vortices, reveals two different behaviors depending on the region considered. Indeed, region 3 is close to the edge of the sample, leading to a larger density of vortices at low field, since they enter by the edge of the specimen and then are pushed towards the middle of the sample \cite{Olsen}. The high value of $\left|\Psi_6\right|$ at low $B$, as well as the shortest distance between vortices, confirm this trend. On the other hand, close to the edge, inhomogeneities are enhanced and deeply affect the movement of the vortices and the arrangements they adopt leading to the flux creep regime in this small portion of the sample (visible at the bottom of region 3). An example of a distorted vortex network at the edge is shown in the bottom right panel of Fig. \ref{triangulation}. When the magnetic field increases, vortices tend to form an Abrikosov lattice in the rest of the specimen, as shown for B = 188.7 G (Fig.~\ref{vortex}), where due to a slightly different specimen orientation (small tilt) the bend contours are shifted and the geometry of the vortex network is revealed. At 117.3 G, the average value of $\left|\Psi_6\right|$ = 0.73, and reaches 0.91 locally (representing around 1/3 of the analyzed region). The theoretical calculation of the average distance for a triangular array (Fig. \ref{six} B)) is in good agreement with the experiments, so validating our calculated magnetic fields. Moreover, the crossing of the dashed curves in the same graph shows that vortices move from one region to another, confirming that bend contours are only related to diffraction contrast and do not affect the geometry of the vortex lattice.

In conclusion, the authors report the imaging of flux vortices in MgB$_2$ single crystals using Cryo-Lorentz Transmission Electron Microscopy. Transmission Electron Microscopy has the advantage of being sensitive to the magnetic field throughout the thickness of the sample and not only at the surface. The vortex lattices exhibit the behavior predicted by the theory \cite{Abrikosov}. No stripes or gossamer-like patterns were observed. Rather, their arrangment tends to that of the hexagonal Abrikosov lattice confirming that MgB$_2$ is a prototypical type-II superconductor. This finding is related to the high surface quality obtained during the sample preparation by focused ion beam that allows the direct observation of the vortex dynamics in a defect-free specimen. If the trend of $\left|\Psi_6\right|$ is to be verified for larger values of $B$ (around 200 G and above) spatial resolution will need to be improved in order to distinguish individual quanta at these high fields. We conclude that the small distortion of the triangular lattice, even if our range of $B$ is far away from $H_{c2}$, is related to impurities coming from the growing process itself, since the surface obtained after the preparation step is defect free.

\begin{acknowledgments}
This work was supported by the European Research Council Grant 258697 "Ultrafast Spectroscopic Electron Diffraction" and the Swiss National Science Founds, Gesuchnummer 200021-140760.
\end{acknowledgments}


\begin{thebibliography}{29}
\bibliographystyle{unsrt}
\expandafter\ifx\csname natexlab\endcsname\relax\def\natexlab#1{#1}\fi
\expandafter\ifx\csname bibnamefont\endcsname\relax
  \def\bibnamefont#1{#1}\fi
\expandafter\ifx\csname bibfnamefont\endcsname\relax
  \def\bibfnamefont#1{#1}\fi
\expandafter\ifx\csname citenamefont\endcsname\relax
  \def\citenamefont#1{#1}\fi
\expandafter\ifx\csname url\endcsname\relax
  \def\url#1{\texttt{#1}}\fi
\expandafter\ifx\csname urlprefix\endcsname\relax\def\urlprefix{URL }\fi
\providecommand{\bibinfo}[2]{#2}
\providecommand{\eprint}[2][]{\url{#2}}
%

\bibitem{Matsuda}
Y. Matsuda, A. Shibata, K. Izawa, H. Ikuta, M. Hasegawa and Y. Kato, {\em Free flux flow resistivity in a strongly overdoped high-${T}_{c}$ cuprate: The purely viscous motion of the vortices in a semiclassical $\textit{d}$-wave superconductor}, \textit{Phys. Rev. B} \textbf{66}, 014527 (2002).

\bibitem{Bardeen}
J. Bardeen and M.J. Stephen, {\em Theory of the Motion of Vortices in Superconductors}, \textit{Phys. Rev.} \textbf{140}, A1197 (1965).

\bibitem{Kim}
Y.B. Kim and M.J. Stephen, {\em Superconductivity}, \textit{edited by R.D. Parks (Marcel Dekker, New York)}, 1107 (1969).

\bibitem{Deaver}
B. S. Deaver and W. M. Fairbank, {\em Experimental Evidence for Quantized Flux in Superconducting Cylinders}, \textit{Phys. Rev. Lett.} \textbf{7}, 43 (1961).

 \bibitem{Abrikosov}
A. A. Abrikosov, {\em On the Magnetic Properties of Superconductors of the Second Group}, \textit{Sov. Phys. JETP} \textbf{5}, 1174 (1957)

\bibitem{Shibata}
A. Shibata, M. Matsumoto, K. Izawa, Y. Matsuda, S. Lee  and S. Tajima, {\em Anomalous flux flow resistivity in the two-gap superconductor MgB$_2$}, \textit{Phys. Rev. B} \textbf{68}, 060501 (2003).

\bibitem{Harada}
K. Harada, T. Matsuda, J. Bonevich, M. Igarashi, S. Kondo, G. Pozzi, U. Kawabe and A. Tonomura, {\em Real-time observation of vortex lattices in a superconductor by electron microscopy}, \textit{Nature} \textbf{360}, 51 (1992).

\bibitem{Chapman}
J.N. Chapman, E.M. Waddell, P.E. Batson and R.P. Ferrier, {\em The Fresnel mode of Lorentz microscopy using a scanning transmission electron microscope}, \textit{Ultramicroscopy} \textbf{4}, 283 (1979).

\bibitem{Budko}
S. L. Bud'ko, G. Lapertot, C. Petrovic, C. E. Cunningham, N. Anderson, and P. C. Canfield, {\em Boron Isotope Effect in Superconducting MgB$_2$}, \textit{Phys. Rev. Lett.} \textbf{86}, 1877 (2001).

\bibitem{LoudonPRB}
J.C. Loudon, C.J. Bowell, N.D. Zhigadlo, J. Karpinski and P.A. Midgley, {\em Magnetic structure of individual flux vortices in superconducting MgB$_{2}$ derived using transmission electron microscopy}, \textit{Phys. Rev. B} \textbf{87}, 144515 (2013).

\bibitem{Manzano}
F. Manzano, A. Carrington, N. E. Hussey, S. Lee, A. Yamamoto, and S. Tajima, {\em Exponential Temperature Dependence of the Penetration Depth in Single Crystal MgB$_2$}, \textit{Phys. Rev. Lett.} \textbf{88}, 047002 (2013).

\bibitem{Moshchalkov1}
V. V. Moshchalkov, M. Menghini, T. Nishio, Q. H. Chen, A.V. Silhanek, V. H. Dao, L. F. Chibotaru, N. D. Zhigadlo and J. Karpinski, {\em Type-1.5 Superconductivity}, \textit{Phys. Rev. Lett.} \textbf{102}, 117001 (2009).

\bibitem{Caplin}
D. Caplin, Y. Bugoslavsky, L. F. Cohen, L. Cowey, J. Driscoll, J. Moore and G. K. Perkins, {\em Critical fields and critical currents in MgB$_2$}, \textit{Supercond. Sci. Technol.} \textbf{16}, 176 (2003)

\bibitem{Gonnelli}
R. S. Gonnelli, D. Daghero, G. A. Ummarino, V. A. Stepanov, J. Jun, S.M. Kazakov, and J. Karpinski, {\em Direct Evidence for Two-Band Superconductivity in MgB$_2$ Single Crystals
from Directional Point-Contact Spectroscopy in Magnetic Fields}, \textit{Phys. Rev. Lett.} \textbf{89}, 247004 (2002)

\bibitem{Gurevich}
A. Gurevich, S. Patnaik, V. Braccini, K. H. Kim, C. Mielke, X. Song, L.D. Cooley, S. D. Bu, D. M. Kim, J. H. Choi, L. J. Belenky, J. Giencke, M. K. Lee, W. Tian, X. Q. Pan, A. Siri, E. E. Hellstrom, C. B. Eom and D. C. Larbalestier, {\em Very high upper critical fields in MgB$_2$ produced by selective tuning of impurity scattering}, \textit{Supercond. Sci. Technol.} \textbf{17}, 278 (2004)
	
\bibitem{Nishio}
T. Nishio, V. H. Dao, Q. Chen, L. F. Chibotaru, K. Kadowaki, and V. V. Moshchalkov, {\em Scanning SQUID microscopy of vortex clusters in multiband superconductors}, \textit{Phys. Rev. B} \textbf{81}, 020506 (2010)
	
\bibitem{Loudon}
J.C. Loudon, C.J. Bowell, N.D. Zhigadlo, J. Karpinski and P.A. Midgley, {\em Imaging flux vortices in MgB$_2$ using transmission electron microscopy}, \textit{Physica C} \textbf{474}, 18 (2012).

\bibitem{Fischer}
M.R. Eskildsen, M. Kugler, G. Levy, S. Tanaka, J. Jun, S.M. Kazakov, J. Karpinski and Ø. Fischer, {\em Vortex lattice imaging in single crystal MgB$_2$ by scanning tunneling spectroscopy}, \textit{Physica C: Superconductivity} \textbf{388}, 143 (2003).

\bibitem{Cribier}
D. Cribier, B. Jacrot, L. Madhav Rao and B. Farnoux, {\em Mise en evidence par diffraction de neutrons d'une structure periodique du champ magnetique dans le niobium supraconducteur}, \textit{Physics Letters} \textbf{9}, 106 (1964).

\bibitem{Kleiner}
W.H. Kleiner, L.M. Roth and S.H. Autler, {\em Bulk Solution of Ginzburg-Landau Equations for Type II Superconductors: Upper Critical Field Region}, \textit{Phys. Rev.} \textbf{133}, A1226 (1964).

\bibitem{Fletter}
A. L. Fetter, P.C. Hohenberg and P. Pincus, {\em Stability of a Lattice of Superfluid Vortices}, \textit{Phys. Rev.} \textbf{147}, 140 (1966).

\bibitem{Tinkham}
M. Tinkham, {\em Introduction to superconductivity}, \textit{Dover publication, inc., Mineola, New York} \textbf{Second edition}, (1975).

\bibitem{Karpinski}
J. Karpinski, S.M. Kazakov, J. Jun, M. Angst, R. Puzniak, A. Wisniewski and P. Bordet, {\em Single crystal growth of MgB$_2$ and thermodynamics of Mg-B-N system at high pressure}, \textit{Physica C} \textbf{385}, 42 (2003).

\bibitem{Olsen}
{\AA}.~A.~F. Olsen, H. Hauglin, T.H. Johansen, P.E. and Goa D. Shantsev, {\em Single vortices observed as they enter NbSe$_{2}$}, \textit{Physica C} \textbf{408}, 537 (2004).

\bibitem{Marco}
M. Cantoni, M. Uchida, Y. Matsui, S. Takekawa, T. Tsuruta and S. Horiuchi, {\em Observation of the interaction of vortices with dislocations in a Nb superconductor by a cryo-Lorentz EM}, \textit{Journal of Electron Microscopy} \textbf{47(5)}, 443 (1998)

\bibitem{Sow}
C.-H. Sow, K. Harada, A. Tonomura, G. Crabtree, and D. G. Grier, {\em Measurement of the Vortex Pair Interaction Potential in a Type-II Superconductor}, \textit{Phys. Rev. Lett.} \textbf{80}, 2693 (1998).

\bibitem{Delaunay}
B. Delaunay, {\em Sur la Sph\`{e}re vide}, \textit{Otdelenie Matematicheskikh/Estestvennykh Nauk} \textbf{7}, 793 (1934)

\bibitem{Fiji}
J. Schindelin, I. Arganda-Carreras, E. Frise, V. Kaynig, M. Longair, T. Pietzsch, S. Preibisch, C. Rueden, S. Saalfeld, B. Schmid, J.-Y. Tinevez, D. J. White, V. Hartenstein, K. Eliceiri, P. Tomancak and A. Cardona, {\em Fiji: an open-source platform for biological-image analysis}, \textit{Nature Methods} \textbf{9}, 676 (2012).

\bibitem{Meisner}
W. Meissner and R. Ochsenfeld, {\em Ein neuer Effekt bei Eintritt der Supraleitfähigkeit}, \textit{Naturwissenschaften} \textbf{21}, 787 (1933).

\bibitem{Auer}
J. Auer and H. Ullmaier, {\em Magnetic Behavior of Type-II Superconductors with Small Ginzburg-Landau Parameters} \textit{Phys. Rev. B.} \textbf{7}, 136 (1973).

\bibitem{Solomon}
P.R. Solomon and Richard E. Harris, {\em Current-Induced Flow of Domains in the Intermediate State of a Superconductor} \textit{Phys. Rev. B.} \textbf{3}, 2969 (1971).

\end{thebibliography}
\end{document}